\documentclass[a4paper,12pt]{article}
\usepackage{enumerate,amsmath,amsfonts,amssymb,amsthm,mathrsfs}
\usepackage{epsfig,subfigure,graphicx,setspace,latexsym,color}
\usepackage{indentfirst,dsfont}
\usepackage{natbib}
\usepackage{setspace}
\usepackage{latexsym}
\usepackage[all]{xy}
\usepackage{graphicx}

\theoremstyle{plain}
\newtheorem{theorem}{Theorem}[section]
\newtheorem{lemma}[theorem]{Lemma}
\newtheorem{cons}[theorem]{Construction}

\theoremstyle{definition}

\theoremstyle{remark}

  \baselineskip 8 mm
\begin{document}
\title{ NP-Completeness of Hamiltonian Cycle Problem on Rooted Directed Path Graphs}
\author{B. S. Panda\thanks{Computer Science and Application Group,
Department of Mathematics, Indian Institute of Technology Delhi,
Hauz Khas, New Delhi 110 016, INDIA, E-mail:
bspanda@maths.iitd.ac.in, dinabandhu.pradhan@mail2.iitd.ac.in}, D.
Pradhan\thanks{This author was supported by Council of Scientific \&
Industrial Research, INDIA.} \\Indian Institute of Technology Delhi}
\date{}
\maketitle

\abstract{ The Hamiltonian cycle problem is to decide whether a
given graph has a  Hamiltonian cycle.  Bertossi and Bonuccelli
(1986, Information Processing Letters, 23, 195-200) proved that the
Hamiltonian Cycle Problem is NP-Complete even for undirected path
graphs and left the Hamiltonian cycle problem open for directed path
graphs. Narasimhan (1989, Information  Processing Letters, 32,
167-170) proved that the Hamiltonian Cycle Problem is NP-Complete
even for directed path graphs and left the Hamiltonian cycle problem
open for rooted directed path graphs. In this paper we  resolve this
open problem by proving that the Hamiltonian Cycle Problem is also
NP-Complete for rooted directed path graphs.
}\\


\section{Introduction}
Let $G=(V,E)$ be a graph. Let $N(x)$ denote the set of all neighbors
of $x$ in $G$. Let $d(x)=|N(x)|$ denote the degree of $x$. A subset
$ S$ of $V$ is called a clique of $G$ if $G[S]$, the subgraph of $G$
induced on $S$, is a complete subgraph of $G$. $S$ is called a
maximal clique if $S$ is a clique but no proper super set of $S$ in
$G$ is a clique in $G$. By the maximum degree of a graph $G$, we
mean the maximum of the degrees of the vertices in $G$. A cycle $C$
of $G$ is called a {\it Hamiltonian cycle} of $G$ if $C$ contains
all the vertices of $G$. The problem of deciding whether a given
graph $G$ has a Hamiltonian cycle is known as {\bf Hamiltonian cycle
problem}. This problem in general graph is well-known to be
NP-Complete \cite{garey}. It is known to be NP-Complete even when
the inputs are restricted to several classes of interesting special
classes of graphs such as {\it planar cubic $3$-connected graphs}
\cite{dawes}, {\it bipartite graphs} \cite{bipartite}, {\it edge
graphs} (line graphs) \cite{booth}, and {\it chordal graphs}
\cite{stewart}. Bertossi and Bonuccelli \cite{bert} proved that the
Hamiltonian Cycle Problem is NP-Complete even for undirected path
graphs. The Hamiltonian cycle problem for directed path graphs and
circular arc graphs were left open by Bertossi and Bonuccelli
\cite{bert}. Narasimhan \cite{giri} later proved that the
Hamiltonian Cycle Problem is NP-Complete even for directed path
graphs. However, the Hamiltonian cycle problem for circular graphs
was solved   by Shih et al.\cite{hsu} in $O(n^2logn)$ time.
%Note that the linear and $O(n^2)$ time Hamiltonian cycle algorithms
%cited as forthcoming in \cite{liang} were flawed).
The Hamiltonian cycle problem on rooted directed path graph was left
open  by Narasimhan \cite{giri}. The Hamiltonian cycle problem on
rooted directed path graph is also mentioned to be open in
(\cite{spinrad}, page $311$).

In this paper we resolve this open problem. In fact, we prove that
the Hamiltonian cycle problem is also NP-Complete for rooted
directed path graphs. However, it is worth mentioning that the
Hamiltonian cycle problem can be solved in polynomial time for {\it
$2$-sep rooted directed path graphs} \cite{panda}, a proper subclass
of rooted directed path graphs.

 The rest of the paper is organized as follows. In Section $2$, we
introduce the rooted directed path graphs and present some results
on this class of graphs. In Section $3$, we prove that the
Hamiltonian cycle problem is NP-Complete for  rooted directed path
graphs. We use the techniques similar to those used in
\cite{bert,giri}. The reduction is carried out from the Hamiltonian
Cycle Problem on bipartite graph with maximum  degree $3$, which was
proved to be NP-Complete by Itai et al. \cite{itai}.

\section{Rooted directed path graphs}

Two paths, say $P_1$ and $P_2$, in a tree $T$ are said to intersect
if $V(P_1) \cap V(P_2) \neq \emptyset$.

Let $\mathscr{F}$ be a finite family of non-empty sets. An
undirected graph $G$ is an intersection graph for $\mathscr{F}$ if
there is a one-to-one correspondence between the vertices of $G$ and
the sets in $\mathscr{F}$ such that two vertices in $G$ are adjacent
 if and only if their corresponding sets have non-empty intersection.
 If $\mathscr{F}$ is a family of paths in an undirected tree $T$,
then $G$ is called an {\it undirected path} graph. If $\mathscr{F}$
is a family of directed paths in a directed tree $T$, i.e., a tree
in which each edge is oriented, then $G$ is called a {\it directed
path} graph. Note that a directed tree may have more than one vertex
of in-degree zero. A rooted directed tree is a directed tree having
exactly one vertex of in-degree zero. If $\mathscr{F}$ is a family
of directed paths in a rooted directed tree $T$, then $G$ is called
a {\it rooted directed path} graph. Undirected path graphs, directed
path graphs, and rooted directed path graphs are also known as {\bf
undirected vertex graphs or UV graphs}, {\bf directed vertex graphs
or DV graphs}, and {\bf rooted directed vertex graphs or RDV
graphs}, respectively (see \cite{monma}).

Note that in the above definition of rooted directed path graph, the
rooted directed tree $T$ is arbitrary. However, Gavril \cite{gav}
characterized the rooted directed path graphs $G$ in terms of a tree
$T$ such that $V(T)$ is the set of all maximal cliques of $G$.

\begin{theorem}[Clique Tree Theorem,\cite{gav,monma}]\label{ctt}

Let $G=(V,A)$ be a graph and let $\mathscr{K}$ be the set of all
maximal cliques of $G$. For each vertex $v\in V$, let
$\mathscr{K}_v$ be the set of cliques of $\mathscr{K}$ containing
the vertex $v$. Then $G$ is a rooted directed path graph if and only
if there exists a rooted directed tree $T$ with the vertex set
$\mathscr{K}$, such that for every $v\in V$, $T(\mathscr{K}_v)$, the
subtree of $T$ induced on $\mathscr{K}_v$, is a directed path in
$T$.

\end{theorem}
The tree $T$ in the above theorem is called the {\bf RDP clique
tree} for $G$.

\section{NP-Completeness }

Consider the following  problem.

\noindent {\bf Problem $\Pi$ :}\\
{\bf Instance} : A bipartite graph $B=(M,N,E)$ having maximum degree $3$.\\
{\bf Question} : Does $B$ contain a Hamiltonian cycle?

It has been shown in \cite{itai} that the  problem $\Pi$ is
NP-Complete.

\begin{lemma}[\bf Itai et. al \cite{itai}]\label{lemma3}
The problem $\Pi$ is NP-Complete.
\end{lemma}

We will show that the following problem $\Pi_1$ is NP-complete.

\noindent {\bf Problem $\Pi_1$ :}\\
{\bf Instance} : A rooted directed path graph $G=(V,A)$.\\
{\bf Question} : Does $G$ contain a Hamiltonian cycle?

The transformation of an instance of problem $\Pi$ to an instance of
the problem $\Pi_1$ is described below.

Let $B(M,N,E)$, a bipartite graph having $n$ vertices with maximum
degree $3$, be an instance of the problem $\Pi$.  Without loss of
generality, we assume that $n=2r, r \geq 2$, the vertex sets $M$ and
$N$ both have $r$ vertices each and that $B$ has no vertex with
degree one, since otherwise $B$ has  no Hamiltonian cycle. Let
$M=\{m_1, m_2,\ldots,m_r\}$ and $N=\{n_1, n_2,\ldots, n_r\}$. We
show how to construct an instance of the problem $\Pi_1$ by showing
how to construct a directed path graph $G(V,A)$ such that $B$ has a
Hamiltonian cycle if and only if $G$ has a Hamiltonian cycle. We
describe $G$ by describing all its maximal cliques. Note that
describing all the maximal cliques of a graph fully defines the
graph itself.

%\noindent {\bf Construction}
\begin{cons}\label{cons1}
\noindent Corresponding to each  vertex $m_i \in M$, $1 \leq i \leq
r$,  construct the clique $K_i=\{X_i \} \cup \{A_{sj}: m_sn_j\in E,
1 \leq s \leq i, 1 \leq j \leq r\}$. Corresponding to each  vertex
$n_j \in N$ with $d(n_j)=3$,  $1 \leq j \leq r$, construct two
cliques $K_j'=\{Y_j\} \cup \{A_{ij}: m_in_j \in E, 1 \leq i \leq
r\}$ and $K_j''=\{Z_j\} \cup \{A_{ij}: m_in_j\in E, 1 \leq i \leq
r\}$. Corresponding to each  vertex $n_j \in N$ with $d(n_j)=2$, $1
\leq j \leq r$,  construct the  clique $K_j'=\{Y_j\} \cup \{A_{ij}:
m_in_j \in E, 1 \leq i \leq r\}$.

%%\hspace{4cm} $K_k=X_k\cup \{A_{ij}: (m_i, n_j)\in E\}, \forall 1\leq
%%i, j\leq n$
%%
%%
%%\noindent Corresponding to the vertex $n_j\in N$ with degree $3$,
%construct cliques
%
%\hspace{4cm} $K_j'=Y_j\cup \{A_{ij}: (m_i, n_j)\in E\}$
%
%\hspace{4cm} $K_j''=Z_j\cup \{A_{ij}: (m_i, n_j)\in E\}$
%
%\noindent Corresponding to the vertex $n_j\in N$ with degree $2$,
%construct cliques
%
%\hspace{4cm} $K_j'=Y_j\cup \{A_{ij}: (m_i, n_j)\in E\}$

Note that the cliques mentioned above are the only maximal cliques
in $G$. Hence it is clear that, $V(G)=\{X_1,X_2,...,X_r\}\cup \{Y_1,
Y_2, ....,Y_r\} \cup \{Z_j:$ $d(n_j)=3\}$ $\cup \{A_{ij}:m_in_j \in
E, 1 \leq i \leq r, 1 \leq j \leq r\}$

\end{cons}

Figure \ref{1} illustrates the construction of the maximal cliques
of $G$ from a given bipartite graph $B$.

We now prove that the resulting graph $G$ is a rooted directed path
graphs.

\begin{lemma}\label{lemma1}
The graph $G$ constructed by Construction \ref{cons1}, is a rooted
directed path graph.

\end{lemma}

\begin{proof}

Let $\mathscr{K}$ be the set of all maximal cliques of $G$. Hence,
 $\mathscr{K}=\{K_1,K_2,\ldots, K_r\} \cup \{K_1',K_2',\ldots,K_r'\} \cup \{K_j'' :
d(n_j)=3, 1 \leq j \leq r \}$.  Let $T=(\mathscr{K}, A)$ be the
directed graph such that $A= \{K_iK_{i+1}, 1 \leq i \leq r-1\} \cup
\{K_rK_j', 1 \leq j \leq r\} \cup \{K_j'K_j'':  d(n_j)=3, 1 \leq j
\leq r\}$. Clearly, $T$ is a rooted directed tree with root $K_1$.
Figure \ref{1} contains a bipartite graph $G$, the set of maximal
cliques of $G$ constructed by Construction \ref{cons1} and an RDP
clique tree $T$ of  $G$ constructed as above.

%and degree(n_jwith vertex set $\mathscr{K}$. Hence $T$ has a vertex
%for each maximal clique of $G$. Let the directed edges of $T$ as
%follows:
%
%\hspace{4cm}$(K_i, K_{i+1}),   \hspace{.8cm}       \forall i, 1\leq
%i\leq n-1,$
%
%\hspace{4cm}$(K_n, K_j'),  \hspace{1cm}      \forall j, 1\leq j\leq
%n,$
%
%\hspace{4cm}$(K_j', K_j''),\hspace{1cm}    \forall j, 1\leq j\leq n,
%$degree$(n_j)=3$.

%\begin{figure}[t]
%\begin{center}
%\scalebox{.60}{\input{fig5.pstex_t}} \caption{The bipartite graph
%$B$, the set of all maximal cliques of $G$ constructed using
%Construction \ref{cons1}, and an RDP clique tree $T$ of the graph
%$G$.}\label{1}
%\end{center}
%\end{figure}

\begin{figure}
%\center{\includegraphics[ width = 16cm]{fig1.eps}}
\centerline{\hbox{\epsfig{figure=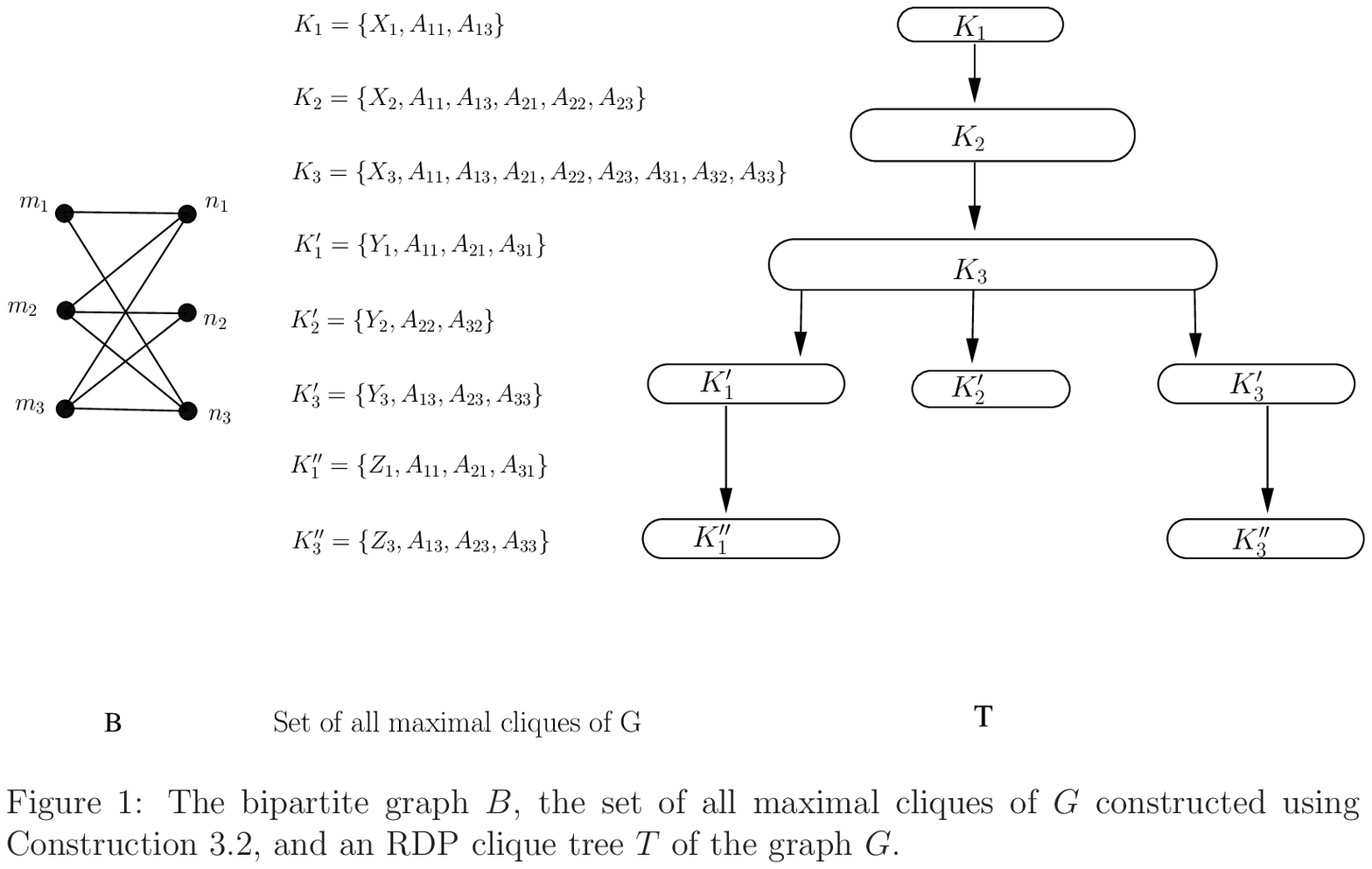,height=10cm,width=14cm}}}
%\vspace{-6cm} \caption{\label{1} The bipartite graph $B$, the set of
%all maximal cliques of $G$ constructed using Construction
%\ref{cons1}, and an RDP clique tree $T$ of the graph $G$.}
\end{figure}

Let $v$ be a vertex of $G$. If $v$ is either $X_i,Y_j$ or $Z_j$,
then $T(\mathscr{K}_v)$ consists of the only one vertex and hence is
a directed path of length zero. If $v=A_{ij}$ and $d(n_j)=3$, then
$T(\mathscr{K}_v)$ consists of the directed path $\langle K_i,
K_{i+1},\ldots, K_r, K_j', K_j'' \rangle$. If $v=A_{ij}$ and
$d(n_j)=2$, then $T(\mathscr{K}_v)$ consists of the directed path
$\langle K_i, K_{i+1},\ldots, K_r,K_j' \rangle$. Hence for each
vertex $v\in V(G)$, $T(\mathscr{K}_v)$ is a directed path in $T$.
Hence by the Theorem \ref {ctt}, $G$ is a rooted directed path
graph.
\end{proof}

\begin{lemma}\label{lemma2}

The bipartite graph $B$ contains a Hamiltonian cycle if and only if
 $G$ contains a Hamiltonian cycle.

\end{lemma}

\begin{proof}
\noindent {\bf Necessity:}

If $B$ has a Hamiltonian cycle $C_B$, we obtain a Hamiltonian cycle
$C_G$ for $G$ as follows. If $m_i, n_j, m_k$ are three consecutive
vertices in $C_B$, we obtain $C_G$ by substituting the sequence
$\langle X_i, A_{ij}, Y_j, A_{hj}, Z_j, A_{kj}, X_k\rangle$ if
$d(n_j)=3$ ( in this case, $m_h$ is the third vertex adjacent to
$n_j$), or with $\langle X_i, A_{ij}, Y_j, A_{kj}, X_k\rangle$ if
$d(n_j)=2$. This results in a Hamiltonian cycle for $G$ since all
the vertices are covered.

\noindent {\bf Sufficiency:}

Let $C_G$ be a Hamiltonian cycle for $G$.

\noindent{\bf Claim :} A sequence of the form $\langle X_s, A_{ij},
Y_j, A_{hj}, Z_j, A_{kj}, X_t\rangle$, where $s\geq i, t\geq k$ if

\hspace{1cm}$d(n_j)=3$, or of the form $\langle X_s, A_{ij}, Y_j,
A_{kj}, X_t\rangle$, where $s\geq i, t\geq k$ if $d(n_j)=2$,

\hspace{1cm}must appear in $C_G$

\noindent{\bf Proof of Claim :} If $d(n_j)=3$, then
$N(Y_j)=N(Z_j)=\{A_{ij}: m_in_j\in E\}=\{A_{ij},A_{hj},A_{kj}\}$.
So, the sequence $\langle A_{ij}, Y_j, A_{hj}, Z_j, A_{kj}\rangle$
must appear in $C_G$. If $d(n_j)=2$, then $A_{ij}, Y_j, A_{kj}$ must
appear in $C_G$ as $A_{ij}$ and $A_{kj}$ are the only two neighbors
of $Y_j$ in $G$. We call such a sequence  a {\it $j$-block}. Now
each $A_{ij}$ such that $m_in_j\in E$ is contained in exactly one
$j$-block. There are exactly $r$ distinct $X_i$'s each appearing in
exactly one clique and there are exactly $r$ $j$-blocks. So, each
$j$-block must appear immediately after an $X_s$ and must appear
immediately before an $X_t$ in $C_G$. If a $j$-block, which starts
with $A_{ij}$ and ends with $A_{kj}$,  appears immediately after
$X_s$ and appears immediately before $X_t$ in $C_G$, then $s\geq i$
and $t\geq k$. So, the claim is proved.

Now we obtain $C_B$ from $C_G$ as follows. If a $j$-block appears
immediately after $X_i$ and appears immediately before $X_k$ in
$C_G$, we obtain $C_B$ by substituting the sequence $\langle X_i,
j$-block,$X_k\rangle$ with the sequence $\langle m_i, n_j,
m_k\rangle$. It is easy  to see that the resulting $C_B$ is a
Hamiltonian cycle for $B$.

\end{proof}
Next we show that the problem $\Pi_1$ is NP-Complete.

\begin{theorem}
The problem $\Pi_1$ is NP-Complete.
\end{theorem}
\begin{proof}
Clearly, the problem $\Pi_1$ is in NP. To show that $\Pi_1$ is
NP-hard, we use a transformation from the problem $\Pi$.

Consider an instance of $\Pi$, i.e., a bipartite graph $B=(M,N,E)$
with maximum degree $3$. Construct the graph $G=(V,A)$ from $B$
using {\bf Construction \ref{cons1}}. It can be easily verified that
the construction of $G$ from $B$ can be done in polynomial time. By
Lemma \ref{lemma1}, $G$ is a rooted directed path graph. Again by
Lemma \ref{lemma2}, there exists a Hamiltonian cycle in $G$ if and
only if there exists a Hamiltonian cycle in $B$. So, $B$ is an `yes'
instance of $\Pi$ if and only if $G$ is an `yes' instance of
$\Pi_1$. Since, by Lemma \ref{lemma3}, $\Pi$ is NP-Complete, $\Pi_1$
is also NP-Complete.

\end{proof}

\section{Conclusion} In this paper, we proved that the problem of
deciding whether a given rooted directed path graph $G$ contains a
Hamiltonian cycle is NP-complete. This was an open problem and was
mentioned in \cite{giri} and in (\cite {spinrad}, page $311$).
%\newpage

 \bibliographystyle{plain}
% \bibliography{NP-RDV}

\end{document}